\documentclass[10pt]{article}
\usepackage{epsfig,amsmath,amsfonts}
\author{D. Galakhov\footnote{e-mail: galakhov@itep.ru}\\ \textit{MIPT and ITEP, Russia, Moscow}}
\title{Self-Interaction and Regularization of Classical Electrodynamics in Higher Dimensions}
\date{}
\newcommand{\mul}{\frac{1}{(2\pi)^{D}}}
\def\be{\begin{eqnarray}}
\def\ee{\end{eqnarray}}
\def\nn{\nonumber}
\def\pr{\partial}
\def\opp{\left(\frac{1}{2(Ru)}\frac{d}{ds}\right)}
\def\op{\left(\frac{1}{2(nu)}\frac{d}{ds}\right)}
\begin{document}
\maketitle \vspace{-6.5 cm}\hfill ITEP/TH-36/07 \vspace{5.5 cm}
\begin{center}
\textbf{Abstract}
\end{center}
The classical electrodynamic system of field and a single point-like source is considered in even-dimensional space-time. The problem of self-interaction is discussed. It is manifestly shown that all singular terms appearing in these equations can be regularized. Relations between formulae for radiation and radiation friction are discussed.
\section{Introduction}
To deal with models involving higher dimensions (see \cite{M&M3} and references therein) requires understanding field theory in higher dimensions and, in particular, solving the problems of radiation and self-interaction. These problems are well-investigated in ordinary flat four-dimensional space-time (see, for example, \cite{Lndfsh,Topt}). At the same time, the radiation theory in higher dimensions is not studied in detail until now. Multi-dimensional radiation is considered in a rather restricted set of works, see \cite{M&M2,links}. The most important fundamental point of the radiation theory is interaction between field and a charged particle inducing this field, or self-interaction. In fact, influence of the field on the particle produces a force similar to friction. This is since the particle moving with acceleration loses energy with the radiated field. Integration over the field degrees of freedom transforms this system into a "quasi-particle" being acted with the radiation friction force on. This leads to an infinite mass of the field. The problem of self-interaction and regularization of corresponding infinities is considered in \cite{Kaz} in six dimensions. A receipt for obtaining the equations of motion based on rather general principles relating radiation and radiation friction is described in \cite{Kos} in six-dimensional space-time. Unfortunately, this technique fails already in eight dimensions, see \cite{M&M1}. Therefore, in order to come to higher dimensions, we apply here the method of expansion to series described in \cite{Kaz}. In the present paper the classical model of electro-dynamics with a single particle is considered in any even-dimensional space-time. It is obtained that all the infinite terms can be regularized. It is shown that a radiation friction force can not be treated as extremum of an ordinary action and its relation to the radiated momentum of accelerated particle is discussed.
\section{Equations of motion}
\subsection{The least action principle and equations of motion} The following action is used for a description of motion of a charged particle in electro-magnetic field in D-dimensional space-time (see
\cite{Lndfsh}) \be \label{action} S=-\sum_{a\subset \Upsilon}
m_{a}\int ds_{a} -\alpha \int F_{\mu\nu}F^{\mu\nu}d^{D}x-\sum_{a\subset \Upsilon}e_{a}\int
A_{\mu}dx^{\mu}_{a}. \ee
Here there are three terms: for charged particles, field and their interaction. The set $\Upsilon$ is a set of charged particles. It can be discrete for point-like sources or continuously distributed for charged medium.
A factor $\alpha$ depends on the chosen scale. It is convenient to choose it different in different cases. The conventional technique allows one to get the following equations for the particle and field:
 \be\label{motforpart}
m_{a}\frac{du^{\mu}}{ds}=e_{a}F^{\mu\nu}u_{\nu};\\
\label{eqmfld}{F^{\mu\nu}}_{,\mu}=\frac{1}{4\alpha}j^{\nu};\\
j(x)=\sum_{a\subset \Upsilon} e_{a}\int \delta(x-z(s))dz(s). \ee
As usual the Lorenz gauge is used. In this case the field equation (\ref{eqmfld}) take the form of the wave equation \be \label{waveeq}\Box A=\frac{1}{4\alpha}j. \ee
We shall use the Green function to construct its solutions.
\subsection{The Green function for the wave equation}
Consider the equation \be \label{Grf}\Box^{(D)}G^{(D)}(x)=\delta(x).
\ee It is well known that the solution of (\ref{waveeq}) is given by the following expression:
\be\label{pot}
 A=\int G(x'-x)\frac{1}{4\alpha}j(x')dx'.
\ee
 For solving (\ref{Grf}), on can make the Fourier transform of the whole equation. The resulting equation reads
  \be
 k^2\tilde{G}(k)=1.
\ee
 One can easily find its solution
\be
 \tilde{G}(k)=\frac{1}{k^2}+\Phi(k)\delta(k^2),
\ee where $\Phi(k)$ is an arbitrary function. The term $\Phi(k)\delta(k^2)$ is responsible for free electro-magnetic field, which is not produced by any source, and will not be considered further. Then the expression for the Green function has the following form
\be
G^{(D)}(x)=\mul\int d^{D}k \frac{e^{i(kx)}}{k^2}
=\mul\int\limits_{-\infty}^{\infty} d\omega \int \frac{e^{i(\omega
t-(kr))}}{\omega^2-k^2}d^{D-1}k. \ee
After quite technical calculation one obtains the relation for the retarded Green function:
 \be
G^{(D)}=\frac{1}{2}\pi^{(2-D)/2}\theta(x_0)\delta^{(D/2-2)}(x^2),{\bf\qquad
for\; even \; D};\\
G^{(D)}=\frac{(-1)^{(D-3)/2}}{2}\pi^{-D/2}\Gamma\left(\frac{D-2}{2}\right)\theta(x_0-|x|)(x^2)^{(2-D)/2},{\bf\qquad
for \; odd}. \ee
One can observe from these formulae that in odd-dimensional space-time the interaction is nonlocal because the whole history of previous motion of the particle affects its present motion, while in even-dimensional space-time the final expression for potential (\ref{pot}) would depend just on higher derivatives of the velocity. This very case is considered further.
\section{"Self-Interaction"}
\subsection{Integration of equations of motion}\label{sbsecint}
To deal with the self-interaction one should consider a system of a point-like source and the field it produces.
It is necessary to solve the problem of how the field of some particle affects itself.
 In order to take into account the influence of the field on the particle, one may follow the simplest way: to substitute solution of the field equation in the equation of motion of the particle. However, one encounters some difficulties on this way: the field is singular in the very particle position. Still, there is a technique that allows one to study these singularities and to exclude them. First, one has to choose the parameter $\alpha=\frac{1}{8}\pi^{(2-D)/2}$ such that the solution of equation (\ref{waveeq}) for a point source becomes of the following form \be\label{sol}
A=e\int d\tilde{s} \delta_{-}^{(D-4)/2}((x(s)-x(\tilde{s}))^2)u(\tilde{s}),
 \ee
where $u$ is D-velocity of the point-like source.
The notation $\delta_{-}$ is introduced here for $\theta(-t)\delta(x)$ where $\theta(x)$ is the Heaviside $\theta$-function. Integration goes over the whole world-line. Then one gets from equation (\ref{motforpart}) the equation for the self-interaction
 \be
 \label{eqmot} m\frac{du}{ds}=2e^2\int d\tilde{s}
  \delta_{-}^{(D-2)/2}((x-\tilde{x})^2)([x-\tilde{x},\tilde{u}],u).
 \ee
 Following the technique introduced in \cite{Topt} and
\cite{Kaz} after change $\tilde{s}=s+\sigma$ one can rewrite equation (\ref{eqmot}) as \be
 \nn m\frac{du}{d s}=2e^2\int d\sigma \delta_{-}^{(D-2)/2}(\sigma^2+((x(s)-x(s+\sigma))^2-\sigma^2))\times\\
 \times([x(s)-x(s+\sigma),u(s+\sigma)],u(s))
 \ee
and expand it to the Taylor series:
 \be\label{Tser} m\frac{du}{ds}=2e^2\int
\nn\sum\limits_{k=0}^{\infty}
\frac{\delta_{-}^{(D-2)/2+k}(\sigma^2)}{k!}\left((x(s)-x(s+\sigma))^2-\sigma^2
\right)^k\times\\
\times([x(s)-x(s+\sigma),u(s+\sigma)],u(s)). \ee
Now it is necessary to compute the following integral
 \be\nn
\mathbb{A}^{m}_{N}=\int\limits_{-\infty}^{\infty}
\delta_{-}^{(N)}(t^2)t^{m}dt, \ee or, rewritten, \be\nn
\mathbb{A}^{m}_{N}=\int\limits^{0}_{-\infty}\delta^{(N)}(t^2)t^{m}dt.
\ee After changing the variable one obtains \be\nn
\mathbb{A}^{m}_{N}=\frac{(-1)^{\left[\frac{m}{2}\right]}}{2}\int\limits_0^{\infty}\delta^{N}(x)x^{(m-1)/2}dx.
\ee The integration at the point 0 is undefined here. We suppose that one integrates from -0. $\mathbb{A}^{m}_{N}$ does not vanish just
only in the case $m\leq 2N$ for even $m$ and $m=2N+1$. Then one finally
gets \be
\mathbb{A}^{2n}_{N}=\frac{(-1)^{n}}{2}\mathcal{C}^{n-1}_{n-1/2}(n-1)!\int\limits_{0}^{\infty}\frac{\delta^{N-n}(x)}{\sqrt{x}}dx.
\ee All of these integrals are singular. However, the following expression is
finite: \be \mathbb{A}^{2N+1}_{N}=\frac{(-1)^{N+1}}{2}N!\ee Using this results one can notice that
the expression (\ref{Tser}) being expanded to the series in $\sigma$ gives
only finite number of nonzero terms. At first, consider the terms
with even powers of $\sigma$. All of them have infinite
coefficients. Note that, as the space-time dimension increases, the structure of terms in the series (velocity polynomials) does not change, just new terms emerge and orders of singularities grows. Therefore, one suffices to consider the only term with $m=2N$ reproducing all necessary structures. It has
the following form \be 2e^2\sum\limits_{p+l+t \leq
D-3+2k}\frac{1}{k!}\mathbb{A}^{D-2+2k}_{(D-2)/2+k}U^{p}_{k}\frac{([u^{(l)},u^{(t)}],u)}{(l+1)!t!},
\ee where the new notation is introduced\be\nn
U^{p}_{k}=\prod\limits_{\alpha=1}^{k}\frac{(u^{(i_{\alpha})}u^{(j_{\alpha})})}{(i_{\alpha}+1)!(j_{\alpha}+1)!}\quad;\qquad\sum_{\alpha=1}^{k}(i_{\alpha}+j_{\alpha})=p\quad;\qquad
i_{\alpha}+j_{\alpha}\geq 2, \forall \alpha. \ee These terms are
singular, however, if they are Lagrangian, i. e. they can be obtained by varying some counter-terms in the action, then one suffices to redefine the action in order to obtain the finite
equations of motion of "dressed" particle.
\subsection{Relation between radiation and radiation friction}
The finite term has the following form \be \label{radfr}
2e^2\sum\limits_{p+l+t \leq
D-2+2k}\frac{(-1)^{D/2+k}((D-2)/2+k)!}{k!}U^{p}_{k}\frac{([u^{(l)},u^{(t)}],u)}{(l+1)!t!}.
\ee This term is expected to be non-Lagrangian because our system
is not closed and (\label{radfr}) can be treated as friction due to
D-momentum losses due to radiation. Calculation of these losses is the
matter of technique.
The energy-momentum tensor
for electro-magnetic field is (see \cite{Lndfsh}) \be
T^{\mu\nu}=4\alpha(-F^{\mu\lambda}{F^{\nu}}_{\lambda}+\frac{1}{4}g^{\mu\nu}F_{\lambda\sigma}F^{\lambda\sigma})
\ee Being integrated over some surface at infinity, this tensor
gives the total radiated D-momentum, i. e. \be \frac{d P^{\mu}}{d
s}=-\oint\limits_{\sigma}T^{\mu\nu}d\sigma_{\nu}. \ee On the other hand, it is
easy to check that the obtained formulae for the radiation friction and
radiated momentum do not coincide even in the simplest case of four dimensions. It
may be treated as if their difference is absorbed into the derivative
of momentum, i. e. this difference is the total derivative of some function.
In this way, one can predict the formula for radiation friction for $D=4,6$. This approach is discussed in 
\cite{Kos,M&M1}, while for $D\geq 8$ it fails, and one has to use another technique.
\subsection{Receipt for the equation of motion from the action}
In subsection \ref{sbsecint} it was mentioned that
non-Lagrangian terms can be regularized by redefining the
action of the particle. One may expect there is a receipt to obtain the
corresponding terms at the level of action (\ref{action}).
For this let us substitute the obtained solution of equation
(\ref{waveeq}) into the action (\ref{action}). Variation of
thus obtained action $S^{*}$ differs from the variation of the old one
because variations of the field and the particle trajectory are not independent, but related by (\ref{waveeq}). On the other hand, this relation implies that the field on the light cone with vertex at the point of the particle trajectory explicitly depends on the motion of the particle in this point only. Therefore we choose to integrate the second integral in (\ref{action}) D-volume to be between the light cones with vertexes in the initial and final points of the particle trajectory. Then variations of the field on these cones and the particle trajectory in the initial and final points are equal to zero simultaneously. Finally, one gets the expression for the redefined action $S^{*}$
\be\nn \delta S^{*}=-m\int \delta ds - e\int \delta
x^{\nu}\pr_{\nu} A_{\mu} dx^{\mu}-e\int A_{\mu}\delta
dx^{\mu}-e\int \delta A_{\mu} dx^{\mu}\\-4\alpha\int
F^{\mu\nu}\pr_{\mu}\delta A_{\nu}d^{D}x. \ee Notice that
$A_{\mu}[x(s)]$ is the functional given by (\ref{sol}). Using the
Gauss theorem and equation (\ref{waveeq}) one can derive
\be\label{viol} \delta S^{*}=\delta S
-4\alpha\int\limits_{\Sigma}F^{\mu\nu}\delta A_{\nu}d\Sigma_{\mu}.
\ee Then the equation given by the variation of $S^{*}$ differs from (\ref{eqmot}) by
the divergence obtained above. This divergence gives non-Lagrangian terms
in the final equation of motion. One suffices to check
that the second variation is nonsymmetric to show this. It means that \be\nn
\delta_{y}\int F^{\mu\nu}\delta_{x} A_{\nu}d\Sigma_{\mu}=\\
= \int \pr^{\mu}\delta_{y}A^{\nu}\delta_{x}A_{\nu}d\Sigma_{\mu}-
\int \pr^{\nu}\delta_{y}A^{\mu}\delta_{x}A_{\nu}d\Sigma_{\mu}+\int
F^{\mu\nu}\delta_{y}\delta_{x} A_{\nu}d\Sigma_{\mu}. \ee Thus a criterion whether this term is Lagrangian is not fulfilled: 
\be\nn \delta_{y}\delta_{x}S-\delta_{x}\delta_{y}S=\int\left(
\delta_{y}F^{\mu\nu}\delta_{x}A_{\nu}-\delta_{x}F^{\mu\nu}\delta_{y}A_{\nu}
\right)d\Sigma_{\mu}=\\
=\int d^{D}x \left((\Box
\delta_{y}A,\delta_{x}A)-(\Box\delta_{x}A,\delta_{y}A) \right)\neq
0. \ee
 It is simple to rewrite $S^{*}$ \be\label{othact} \nn
S^{*}=-m\int ds -e\int A_{\mu}dx^{\mu} -2\alpha \int
F^{\mu\nu}\pr_{\mu}A_{\nu}d^{D}x=\\
\nn= -m\int ds -e\int A_{\mu}dx^{\mu}+2\alpha\int A_{\mu}\Box
A^{\mu}d^{D}x-2\alpha\int\limits_{\Sigma}F^{\mu\nu}A_{\nu}d\Sigma_{\mu}=\\
=-m\int ds -\frac{e}{2}\int
A_{\mu}dx^{\mu}-2\alpha\int\limits_{\Sigma}F^{\mu\nu}A_{\nu}d\Sigma_{\mu}
\ee
First consider the second term in the final expression
 \be -\frac{e^2}{2}\int\int
\delta^{((D-4)/2)}((x-\tilde{x})^2)(dxd\tilde{x})
\ee
It can be regularized with application of the same technique which was used for the equation of motion \be
\frac{e^2}{2}\int ds\int d\sigma\sum\limits_{k=0}^{\infty}
\frac{\delta^{(D-2)/2+k}(\sigma^2)}{k!}\left((x(s)-x(s+\sigma))^2-\sigma^2
\right)^k(u(s)u(s+\sigma))
 \ee
Its variation gives \be
-e^2\int\delta^{((D-2)/2)}((x(s)-x(\tilde{s}))^2)([x(s)-x(\tilde{s}),u(\tilde{s})],u(s))d\tilde{s} \ee Note
that the symmetric part (with even $\sigma$ in (\ref{Tser})) of $\delta_{-}$ is $\frac{1}{2}\delta$, the
coefficient emerges since the integration goes only over the negative part of the real axis, then, the
antisymmetrical parts are excluded due to the symmetry of $\delta^{(N)}(\sigma^2)$. It means that the second term
of $S^{*}$ contains all the infinite Lagrangian terms. As it was said previously, it means that they can be regularized
or excluded at all. The former means appearance of additional "masses". Being Lagrangian, these
terms imply that they are the total derivatives and can be treated as contributing to of D-momentum. Calculation of
the final term in (\ref{othact}) gives (see Appendix) \be -2\alpha e \int
\frac{d}{ds}\left(\op^{(D-4)/2}\frac{u}{2(nu)}\right)^2d\Omega ds. \ee This expression can be integrated over
and excluded from the action. Then it is clear that the divergency given by (\ref{viol}) is responsible for the
radiation friction force. Consider the variation of the field caused by the variation of the particle trajectory.
 Using the formula for divergency (\ref{viol}), one gets the expression for the radiation friction force
 \be
 f_{\mu}=4\alpha e^2(-1)^{(D-2)/2}\int
 d\Omega_{D-1}\left\{ \frac{n_{\mu}u_{\nu}}{(nu)}-\eta_{\mu\nu}
 \right\}\frac{d}{ds}\op^{D-3}\frac{u^{\nu}}{2(nu)}.
 \ee
 This formula is not as convenient for technical computations as (\ref{radfr}), on the other hand, the very procedure of the derivation shows that the radiation friction can not be obtained as an extremal of action, and it allows one to check the relation between the radiation and the friction. The derivative of D-momentum obtained with integration of the energy-stress tensor is
 \be
 f^{(rad)}_{\mu}=4\alpha e^2\int d\Omega_{D-1}\frac{n_{\mu}}{(nu)}\left\{\frac{d}{ds}\op^{(D-4)/2}\frac{u}{2(nu)}\right\}^2
 \ee
 Using the Leibnitz rule one can show that these two expressions differ just by a total derivative as it is expected.
\section{Conclusion}
The proposed procedure of integration over field freedom degrees in the action shows in general that all the divergencies are Lagrangian, the radiation friction is well-defined, and one can get the finite equation of motion for a particle with the friction force. However, the precise expression for these divergencies strongly depends on the regularization procedure. One might expect an appearance of all possible Lagrangian terms restricted just by the Poincar\'e invariance, i. e. one suffices to consider in the action the terms made from products of $(u^{(i)}u^{(k)}) $ and some coefficients, which may be infinite. One could also expect that the terms with the same order singularity have to have the same total degree of derivatives. It is since the diverging quantity is scaled with a regularizing length that can be treated, for example, as a size of the particle. On the other hand, one can hardly a priori say anything concrete about relations between coefficients of these terms. However, such relations are obtained, despite the proposed procedure of regularization does not define a concrete representation for the delta-function, i. e. it presents a rather wide class of regularizations. An existence of these relations can be induced by some symmetry. However, such symmetry has to have a rather general character, since the number of the possible additional term combinations grows as their order increases.
\section*{Acknowledges}

I would like to thank A. Morozov and A. Mironov for an informative discussion. This work is partly supported by the Federal Agency for Atomic Energy of Russia, by the Russian President's Grant of Support for the Scientific Schools NSh-8004.2006.2, by RFBR grant 07-02-00878.
\section*{Appendix}

\subsection*{Calculation of surface integrals}
\label{app} To calculate integrals similar to the final term of
(\ref{othact}), the technique introduced in \cite{M&M2}. One fixes some
coordinate frame to compute the third term and chooses the surface
$\Sigma$ given by a light-like vector $R$ originated from every
point of the trajectory. Then \be\nn
d\Sigma^{\mu}=-\epsilon^{\mu\nu\gamma_1\ldots\gamma_{D-2}}u_{\nu}\frac{\pr
R_{\gamma_1}}{\pr \theta_1}\ldots\frac{\pr R_{\gamma_{D-2}}}{\pr
\theta_{D-2}}dsd\theta_1\ldots d\theta_{D-2}, \ee where
$R^{\mu}=R(1,\cos\theta_1,\sin\theta_1\cos\theta_2,\ldots)=Rn^{\mu}$.
In particular, \be \pr_{\mu}s=\frac{R_{\mu}}{(Ru)}. \ee Taking derivatives
of $R$ decreases its power. Because it is assumed that
$R\rightarrow\infty$ then $\frac{dR}{ds}=0$. It is convenient to
represent this expression in terms of $R$ and $u$ \be\nn A=e\int
\delta_{-}^{((D-4)/2)}((x-\tilde{x})^2)\tilde{u}\frac{d(x-\tilde{x})^2}{2(\tilde{x}-x,\tilde{u})}=\\
\nn
=-e\left(-\frac{d\tilde{s}}{2(x-\tilde{x})^2}\frac{d}{d\tilde{s}}\right)^{(D-4)/2}\frac{\tilde{u}}{2(\tilde{x}-x,\tilde{u})}=\\
=e\opp^{(D-4)/2}\frac{u}{2(Ru)}. \ee For the variation
 \be
 \nn\delta A =2\int
 \delta_{-}^{((D-2)/2)}((x-\tilde{x})^2)(\tilde{x}-x,\delta\tilde{x})d\tilde{x}+\int\delta_{-}^{((D-4)/2)}((x-\tilde{x})^2)d\delta\tilde{x}=\\
 =-\opp^{(D-2)/2}\frac{(R\delta x)u}{(Ru)}+\opp^{(D-2)/2}\delta x.
 \ee Then, the final term of (\ref{othact}) is equal to
\be\nn -2\alpha e^2\int\limits_{\Sigma}\left(
\frac{R^{\mu}}{(Ru)}\frac{d}{ds}\opp^{(D-4)/2}\frac{u^{\nu}}{(Ru)}-\frac{R^{\nu}}{(Ru)}\frac{d}{ds}\opp^{(D-4)/2}\frac{u^{\mu}}{(Ru)}
\right)\times\\
\times\opp^{(D-4)/2}\frac{u_{\nu}}{2(Ru)}d\Sigma_{\mu}. \ee
 Since the vector $R^{\mu}$ can be brought through
the derivatives the second term in the brackets can be excluded \be -\alpha e \int
\frac{d}{ds}\left(\op^{(D-4)/2}\frac{u}{2(nu)}\right)^2d\Omega ds, \ee where \be
d\Omega=-\frac{1}{(nu)}\epsilon^{\mu\nu\gamma_1\ldots \gamma_{D-2}}n_{\mu}u_{\nu}\frac{\pr n_{\gamma_1}}{\pr
\theta_1}\ldots\frac{\pr n_{\gamma_{D-2}}}{\pr \gamma_{D-2}}\prod\limits_{i=1}^{D-2}d\theta_{i}.
 \ee
For the sake of convenience, we choose the velocity to be directed along the first axis:
 \be\nn
 d\Omega=\frac{\sqrt{1-v^2}}{1-v\cos\theta_1}\left|
 \begin{array}{cccc}
 1&\cos\theta_1&\sin\theta_1\cos\theta_2&\ldots\\
 \frac{1}{\sqrt{1-v^2}}&\frac{v}{\sqrt{1-v^2}}&0&\ldots\\
 0&-\sin\theta_1&\cos\theta_1\cos\theta_2&\ldots\\
 \ldots
 \end{array}
 \right|\prod\limits_{i=1}^{D-2}d\theta_{i}=\\
 \nn=\frac{1}{1-v\cos\theta_1}\left(\frac{1}{\cos\theta_1}\left|
 \begin{array}{cc}
 \cos\theta_1\cos\theta_2&\ldots\\
 \ldots
 \end{array}
 \right|-
 v\left|
 \begin{array}{cc}
 \cos\theta_1\cos\theta_2&\ldots\\
 \ldots
 \end{array}
 \right|
 \right)\prod\limits_{i=1}^{D-2}d\theta_{i}=\\
 =d\Omega_{D-1}
 \ee
It means that this expression coincides with an infinitesimal element of
(D-1)-solid angle and doesn't depend on the proper time of the
particle, can be brought through derivatives. Calculating such integrals requires calculating of the following integral tensors
\be
\int\frac{n^{\mu_1}n^{\mu_2}\ldots n^{\mu_{s}}}{(nu)^{D-2+s}}d\Omega_{D-1}.
\ee
To calculate them, one can use a general formula introduced in \cite{M&M2} or a recurrent relation:
\be
\int\frac{n^{\mu_1}n^{\mu_2}\ldots n^{\mu_{s+1}}}{(nu)^{D-2+s}}d\Omega=\left\{u^{\mu_{s+1}}-\frac{1}{D-2+s}\left(\eta^{\mu_{s+1}\nu}-u^{\mu_{s+1}}u^{\nu}\right)\frac{\pr}{\pr u^{\nu}}\right\}\int\frac{n^{\mu_1}n^{\mu_2}\ldots n^{\mu_{s}}}{(nu)^{D-2+s}}d\Omega_{D-1}.
\ee

 \subsection*{Examples}
 Several examples are presented here. There are formulas for radiation friction in different dimensions in this table. All these formulae can be easily computed with the help of a program for symbolic calculations. In this table, one must multiply every expression by $e^2$

\bigskip

 \begin{tabular}{|c|c|c|c|c|}
   \hline
   Dim & Rad. friction  \\
   \hline
   &\\
   4 & $\frac{2}{3}(\ddot{u}+{\dot{u}}^2u)$\\
   &\\
   6 & $-\frac{2}{15}(\dot{u}u^{(3)})u-\frac{1}{10}\ddot{u}^2u-\frac{1}{30}u^{(4)}-\frac{1}{4}(\dot{u}\ddot{u})\dot{u}-\frac{1}{12}\dot{u}^4u-\frac{1}{12}\dot{u}^2\ddot{u}$ \\
   &\\
   8 &
   $\frac{7}{60}\dot{u}^2
   (\dot{u}u^{(3)})u+\frac{1}{560}u^{(6)}+\frac{5}{24}\dot{u}^2(\dot{u}\ddot{u})\dot{u}+\frac{5}{144}\dot{u}^4\ddot{u}+\frac{5}{144}\dot{u}^6u+\frac{3}{16}(\dot{u}\ddot{u})^2u+\frac{1}{80}\dot{u}^2u^{(4)}+$\\&$+\frac{3}{280}(\dot{u}u^{(5)})u+\frac{3}{112}(\ddot{u}u^{(4)})u+\frac{1}{56}(u^{(3)})^2u+\frac{3}{80}(\dot{u}u^{(4)})\dot{u}+\frac{1}{12}(\ddot{u}u^{(3)})\dot{u}+\frac{17}{180}\dot{u}^2\ddot{u}^2u+\frac{41}{720}\dot{u}^2\ddot{u}+$\\&$+\frac{1}{16}(\dot{u}\ddot{u})u^{(3)}+\frac{1}{15}(\dot{u}u^{(3)})\ddot{u}$ \\
   &\\

   \hline

 \end{tabular}

\bigskip

New additional Lagrangian terms emerge as the dimension grows. In the table, only the new terms in each dimension are written.

\bigskip

\begin{tabular}{|c|c|c|}
  \hline
  Dim & Lagrangian terms & Variations of these terms\\
  \hline
   && \\
  4 & 1 & $\dot{u}$ \\
  &&\\
  6 & $\dot{u}^2$ & $\frac{3}{2}\dot{u}^2\dot{u}+3(\dot{u}\ddot{u})u+u^{(3)}$\\
  &&\\
  8&1$5(\dot{u}u^{(3)})+10\ddot{u}^2+\frac{85}{8}\dot{u}^4$ & $-\frac{135}{4}(\dot{u}u^{(3)})\dot{u}-\frac{55}{2}\ddot{u}^2\dot{u}-\frac{595}{32}\dot{u}^4\dot{u}-\frac{595}{8}(\dot{u}\ddot{u})\dot{u}^2u$\\
  &&$ -\frac{85}{2}(\dot{u}\ddot{u})\ddot{u}-\frac{85}{8}\dot{u}^2u^{(3)}-\frac{25}{2}(\dot{u}u^{(4)})u-25(\ddot{u}u^{(3)})u-\frac{5}{2}u^{(5)}$\\
  &&\\
  \hline
\end{tabular}

\bigskip

  This table represents the terms whose derivative is the difference between the radiation friction force and the radiated power. For using this one must multiply every expression by $e^2$

\bigskip

\begin{tabular}{|c|c|}
  \hline
  Dim & terms \\
   \hline
   &\\
  4 & $\frac{2}{3}\dot{u}$ \\
  &\\
  6 & $-\frac{8}{105}\dot{u}^2\dot{u}-\frac{2}{15}(\dot{u}\ddot{u})u-\frac{1}{30}u^{(3)}$\\
  &\\
  8 & $ \frac{1}{560}u^{(5)}+\frac{403}{3780}\dot{u}^2(\dot{u}\ddot{u})u+\frac{577}{15120}(\dot{u}\ddot{u})\ddot{u}+\frac{4841}{166320}\dot{u}^4\dot{u}+\frac{1}{80}\dot{u}^2u^{(3)}+\frac{71}{3780}\ddot{u}^2\dot{u}+\frac{3}{280}(\dot{u}u^{(4)})u+$\\&$+\frac{9}{560}(\ddot{u}u^{(3)})u+\frac{3}{112}(\dot{u}u^{(3)})u$\\
  &\\
  \hline
\end{tabular}

\bigskip

One can compare formulas for the radiated power and the radiation friction. Formulas for the radiated power are presented in \cite{M&M2}, several formulas are also presented here, but with the
$\alpha$-factor introduced above.

\bigskip

\begin{tabular}{|c|c|c|c|c|}
   \hline
   Dim & Rad. momentum  \\
   \hline
   &\\
   4 & $\frac{2}{3}{\dot{u}}^2u$\\
   &\\
   6 & $-\frac{1}{12}\dot{u}^4u+\frac{1}{30}\ddot{u}^2u+\frac{1}{28}(\dot{u}\ddot{u})\dot{u}-\frac{1}{140}\dot{u}^2\ddot{u}$ \\
   &\\
   8 &
   $\frac{1}{560}(u^{(3)})^2u+\frac{19}{1890}\dot{u}^2(\dot{u}u^{(3)})u-\frac{389}{15120}(\dot{u}\ddot{u})^2u-\frac{23}{
1890}\dot{u}^2\ddot{u}^2u+\frac{5}{144}\dot{u}^6u+\frac{11}{3780}(\ddot{u}u^{(3)})\dot{u}-$\\&$-\frac{1223}{83160}\dot{u}^2(\dot{u}\ddot{u})\dot{u}+\frac{13}{7560}(\dot{u}u^{(3)})\ddot{u}+\frac{467}{83160}\dot{u}^4\ddot{u}-\frac{1}{1512}(\dot{u}\ddot{u})u^{(3)}
$ \\
   &\\

   \hline

 \end{tabular}

\end{document}